\documentclass{emulateapj}

\usepackage{graphicx}
\usepackage{float}
\usepackage{amsmath}
\usepackage{epsfig,floatflt}

\begin{document}

\title{Increasing evidence for hemispherical power asymmetry in the
  five-year WMAP data}

\author{J. Hoftuft\altaffilmark{1}, H.\ K.\ Eriksen\altaffilmark{1,2},
  A. J.  Banday\altaffilmark{3,4}, K.\ M.\ G\'orski\altaffilmark{5,6,7},
  F. K.  Hansen\altaffilmark{1} and P.\ B.\ Lilje\altaffilmark{1,2}}

\email{h.k.k.eriksen@astro.uio.no}

\altaffiltext{1}{Institute of Theoretical Astrophysics, University of
  Oslo, P.O.\ Box 1029 Blindern, N-0315 Oslo, Norway}

\altaffiltext{2}{Centre of Mathematics for Applications, University of
  Oslo, P.O.\ Box 1053 Blindern, N-0316 Oslo, Norway}

\altaffiltext{3}{Centre d'Etude Spatiale des Rayonnements, 9, av du
  Colonel Roche, BP 44346, 31028 Toulouse Cedex 4, France}

\altaffiltext{4}{Max-Planck-Institut f\"ur Astrophysik,
  Karl-Schwarzschild-Str.\ 1, Postfach 1317, D-85741 Garching bei
  M\"unchen, Germany}

\altaffiltext{5}{Jet Propulsion Laboratory, 4800 Oak Grove Drive,
  Pasadena CA 91109}

\altaffiltext{6}{California Institute of Technology, Pasadena, CA
  91125}

\altaffiltext{7}{Warsaw University Observatory, Aleje Ujazdowskie 4,
  00-478 Warszawa, Poland}


\begin{abstract}
  Motivated by the recent results of Hansen et al.\ (2008) concerning
  a noticeable hemispherical power asymmetry in the WMAP data on small
  angular scales, we revisit the dipole modulated signal model
  introduced by Gordon et al. (2005). This model assumes that the true
  CMB signal consists of a Gaussian isotropic random field modulated
  by a dipole, and is characterized by an overall modulation
  amplitude, $A$, and a preferred direction, $\hat{p}$. Previous
  analyses of this model has been restricted to very low resolution
  (ie., $3.6^{\circ}$ pixels, a smoothing scale of $9^{\circ}$ FWHM
  and $\ell \lesssim 40$) due to computational cost. In this paper, we
  double the angular resolution (ie., $1.8^{\circ}$ pixels and
  $4.5^{\circ}$ FWHM smoothing scale), and compute the full
  corresponding posterior distribution for the 5-year WMAP data. The
  results from our analysis are the following: The best-fit modulation
  amplitude for $\ell \le 64$ and the ILC data with the WMAP KQ85 sky
  cut is $A=0.072\pm0.022$, non-zero at $3.3\sigma$, and the preferred
  direction points toward Galactic coordinates $(l,b) = (224^{\circ},
  -22^{\circ}) \pm 24^{\circ}$. The corresponding results for $\ell
  \lesssim 40$ from earlier analyses was $A = 0.11\pm0.04$ and $(l,b)
  = (225^{\circ},-27^{\circ})$. The statistical significance of a
  non-zero amplitude thus increases from $2.8\sigma$ to $3.3\sigma$
  when increasing $\ell_{\textrm{max}}$ from 40 to 64, and all results
  are consistent to within $1\sigma$. Similarly, the Bayesian
  log-evidence difference with respect to the isotropic model
  increases from $\Delta \ln E = 1.8$ to $\Delta \ln E = 2.6$, ranking
  as ``strong evidence'' on the Jeffreys' scale. The raw best-fit
  log-likelihood difference increases from $\Delta \ln \mathcal{L} =
  6.1$ to $\Delta \ln \mathcal{L} = 7.3$. Similar, and often slightly
  stronger, results are found for other data combinations. Thus, we
  find that the evidence for a dipole power distribution in the WMAP
  data increases with $\ell$ in the 5-year WMAP data set, in agreement
  with the reports of Hansen et al.\ (2008).
\end{abstract}

\section{Introduction}
\label{sec:introduction}

The question of statistical isotropy in the cosmic microwave
background (CMB) has received much attention within the cosmological
community ever since the release of the first-year Wilkinson Microwave
Anisotropy Probe (WMAP; Bennett et al.\ 2003a) in 2003. The reasons for
this are two-fold. On the one hand, the current cosmological
concordance model is based on the concept of inflation
\citep{starobinsky:1980, guth:1981,linde:1982, muhkanov:1981, starobinsky:1982,
  linde:1983, linde:1994, smoot:1992, ruhl:2003, runyan:2003,
  scott:2003}, which predicts a statistically homogeneous and
isotropic universe. Since inflation has proved highly successful in
describing a host of cosmological probes, most importantly the CMB and
large-scale power spectra, this undeniably imposes a strong
theoretical prior towards isotropy and homogeneity.

On the other hand, many detailed studies of the WMAP sky maps,
employing higher-order statistics, have revealed strong hints of both
violation of statistical isotropy and non-Gaussianity. Some early
notable examples include unexpected low-$\ell$ correlations \citep{de
  Oliveira-Costa:2004}, a peculiar large cold spot in the southern
Galactic hemisphere \citep{vielva:2004}, and a dipolar distribution of
large-scale power \citep{eriksen:2004a}. Today, the literature on
non-Gaussianity and violation of statistical isotropy in the WMAP data
has grown very large, indeed \citep[e.g.,][]{bernui:2006, bielewicz:2005,copi:2006,
  cruz:2005, cruz:2006, eriksen:2004b, eriksen:2004c, eriksen:2005,
  jaffe:2005, jaffe:2006, martinez-gonzalez:2006, mcewen:2008,
  rath:2007, yadav:2008}, and it would be unwise not to consider these
issues very seriously.

Of particular interest to us is the question of hemispherical
distribution of power in the WMAP data, first reported by
\citet{eriksen:2004a} and later confirmed by, e.g.,
\citet{hansen:2004} and \citet{eriksen:2005}. The most recent works on
this topic include those presented by \citet{hansen:2008}, who found
that the power asymmetry extends to much smaller scales than
previously thought, and by \citet{eriksen:2007b}, who quantified the
large-scale power asymmetry in the 3-year WMAP data using an optimal
Bayesian framework.

A separate, but possibly physically related, line of work was recently
presented by \citet{groeneboom:2009}, who considered the specific
model for violation of Lorenz invariance in the early universe,
proposed by \citet{ackerman:2007}. This model involves CMB
correlations with a quadrupolar distribution on the sky, and is thus
orthogonal to the current dipolar model. Surprisingly, when analyzing
the 5-year WMAP data, \citet{groeneboom:2009} found supportive
evidence for this model at the $3.8\sigma$ significance level, when
considering angular scales up to $\ell \le 400$. Thus, assuming that
the WMAP observations are free of unknown systematics, there appears
to be increasing evidence for both dipolar and quadrupolar structure
in the CMB power distribution, at all angular scales.

In this paper, we repeat the Bayesian analysis of
\citet{eriksen:2007b}, but double the angular resolution of the
data. Nevertheless, we are still limited to relatively low angular
resolutions, since the method inherently relies on brute-force
evaluation of a pixel-based likelihood, and therefore scales as
$\mathcal{O}(N_{\textrm{pix}}^3)$. Yet, simply by spending more
computer resources we are able to increase the pixel resolution from
$N_{\textrm{side}}=16$ to 32 and decrease the degradation smoothing
scale from $9^{\circ}$ to $4.5^{\circ}$ FWHM. This provides additional
support for multipoles between $\ell \approx 40$ and 80. While not
sufficient to provide a full and direct comparison with the results of
\citet{hansen:2008}, it is a significant improvement over the results
presented by \citet{eriksen:2007b}.

\section{Overview of model and algorithms}
\label{sec:analysis}

The Bayesian analysis framework used in this paper are very similar to
that employed by \citet{eriksen:2007b}. We therefore only give a brief
overview of its main features here, and refer the reader interested in
the full details to the original paper and references therein.

\subsection{Data model and likelihood}

The starting point for our analysis is the phenomenological CMB signal
model first proposed by \citet{gordon:2005},
\begin{equation}
  \mathbf{d}(\hat{n}) = [1 + f(\hat{n})] \mathbf{s}(\hat{n})
  + \mathbf{n}(\hat{n}).
\label{eq:data_model}
\end{equation}
Here $\mathbf{d}(\hat{n})$ denotes the observed data in direction
$\hat{n}$, $\mathbf{s}(\hat{n})$ is an intrinsically isotropic
and Gaussian random field with power spectrum $C_{\ell}$,
$f(\hat{n})$ is an auxiliary modulating field, and
$\mathbf{n}(\hat{n})$ denotes instrumental noise. 

Obviously, if $f = 0$, one recovers the standard isotropic
model. However, we are interested in a possible hemispherical
asymmetry, and we therefore parametrize the modulation field in terms
of a dipole with a free amplitude $A$ and a preferred direction
$\hat{p}$,
\begin{equation}
f(\hat{n}) = A\,(\hat{n}\cdot\hat{p}).
\end{equation}
The modulated signal component is thus an anisotropic, but still
Gaussian, random field, with covariance matrix
\begin{equation}
\mathbf{S}_{\textrm{mod}}(\hat{n}, \hat{m}) = [1+A\,(\hat{n}\cdot\hat{p})]
\mathbf{S}_{\textrm{iso}}(\hat{n}, \hat{m})[1+A\,(\hat{m}\cdot\hat{p})],
\end{equation}
where
\begin{equation}
\mathbf{S}_{\textrm{iso}}(\hat{n}, \hat{m}) = \frac{1}{4\pi}\sum_{\ell}
(2\ell+1) C_{\ell} P_{\ell}(\hat{n}\cdot\hat{m}).
\end{equation}

We now introduce one new feature compared to the analysis of
\citet{eriksen:2007b}, for two reasons. First, we are interested in
studying the behaviour of the modulation field as a function of
$\ell$-range, and therefore want a mechanism to restrict the impact of
the modulation parameters in harmonic space. Second, we also want to
minimize the impact of the arbitrary regularization noise (see Section
\ref{sec:data}) on the modulation parameters at high
$\ell$'s. Therefore, we split the signal covariance matrix into two
parts, one modulated low-$\ell$ part and one isotropic high-$\ell$
part,
\begin{equation}
\mathbf{S}_{\textrm{total}}= \mathbf{S}_{\textrm{mod}}+\mathbf{S}_{\textrm{iso}},
\end{equation}
where only multipoles between $2\le \ell < \ell_{\textrm{mod}}$ are
included in $\mathbf{S}_{\textrm{mod}}$, and only multipoles at $\ell
\ge \ell_{\textrm{mod}}$ are included in
$\mathbf{S}_{\textrm{iso}}$. (Note that we are not proposing a
physical mechanism for generating the modulation field in this paper,
but only attempt to characterize its properties. This split may or may
not be physically well-motivated, but it does serve a useful purpose
in the present paper as it allows us to study the scale dependence of
the modulation field in a controlled manner.)

Including instrumental noise and possible foreground contamination,
the full data covariance matrix reads
\begin{equation}
\mathbf{C} =
\mathbf{S}_{\textrm{mod}}(A,\hat{p}) + \mathbf{S}_{\textrm{iso}} + \mathbf{N} + \mathbf{F}.
\end{equation}
The noise and foreground covariance matrices depend on the data
processing, and will be described in greater detail in \S \ref{sec:data}. 

We also have to parametrize the power spectrum for the underlying
isotropic component, $C_{\ell}$. Following \citet{eriksen:2007b}, we
choose a simple two-parameter model with a free amplitude $q$ and tilt
$n$ for this purpose,
\begin{equation}
C_{\ell} = q \left(\frac{\ell}{\ell_0}\right)^{n} C_{\ell}^{\textrm{fid}}.
\end{equation}
Here $\ell_0$ is a pivot multipole and $C_{\ell}^{\textrm{fid}}$ is a
fiducial model, in the following chosen to be the best-fit
$\Lambda$CDM power law spectrum of \citet{komatsu:2009}.

Since both the signal and noise are assumed to be Gaussian, the
log-likelihood now reads
\begin{equation}
  -2\log \mathcal{L}(A, \hat{p}, q, n) = \mathbf{d}^T \mathbf{C}^{-1} \mathbf{d} + \log |\mathbf{C}|,
\end{equation}
up to an irrelevant constant, with $\mathbf{C}=\mathbf{C}(A, \hat{p},
q, n)$. 

\subsection{The posterior distribution and Bayesian evidence}

The posterior distribution for our model is given by Bayes' theorem,
\begin{equation}
P(q, n, A, \hat{p} | \mathbf{d}, H) = \frac{\mathcal{L}(q, n, A,
  \hat{p}) P(q, n, A, \hat{p}|H)}{P(\mathbf{d}|H)}.
\end{equation}
Here $P(q, n, A, \hat{p}|H)$ is a prior, and
$P(\mathbf{d|H})$ is a normalization factor often called the
``Bayesian evidence''. Note that we now have included an explicit
reference to the hypothesis (or model), $H$, in all factors, as we
will in the following compare two different hypotheses, namely ``H1:
The universe is isotropic ($A=0$)'' versus ``H2: The universe is
anisotropic ($A\ne0$)''. 

We adopt uniform priors for all priors in the following. Specifically,
we adopt$P(q) = \textrm{Uniform}[0.5, 1.5]$ and $P(n) =
\textrm{Uniform}[-0.5, 0.5]$ for the power spectrum, and a uniform
prior over the sphere for the preferred axis, $\hat{p}$. The
modulation amplitude prior is chosen uniform over $[0,
A_{\textrm{max}}]$, where $A_{\textrm{max}}=0.15$ is sufficiently
large to fully encompass the non-zero parts of the likelihood. If more
liberal priors are desired, the interested reader can easily calculate
the corresponding evidence by subtracting the logarithm of the volume
expansion factor from the results quoted in this paper.

With these definitions and priors, the posterior distribution, $P(q,
n, A, \hat{\mathbf{p}} | \mathbf{d}, H)$ is mapped out with a standard
MCMC sampler. The Bayesian evidence, $E = P(\mathbf{d}|H)$ is computed
with the ``nested sampling'' algorithm \citep{skilling:2004,
  mukherjee:2006}. For further details on both procedures, we refer
the interested reader to \citet{eriksen:2007b}.

For easy reference, we recall Jeffreys' interpretational scale for the
Bayesian evidence \citep{jeffreys:1961}: A value of $\Delta \ln E < 1$
indicates a result ``not worth more than a bare mentioning''; a value
of $1 < \Delta \ln E < 2.5$ is considered as ``significant'' evidence;
a value of $2.5 < \Delta \ln E < 5$ is considered ``strong to very
strong''; and $\Delta \ln E > 5$ ranks as ``decisive''. 

\begin{deluxetable*}{llcccccc}
\tabletypesize{\small}
\tablecaption{\label{tab:results}Summary statistics for modulated CMB
  model posteriors}
\tablecomments{Listed quantities are (\emph{first column}) data
  set (\emph{first column});
  mask (\emph{second column}); maximum multipole used for modulation
  covariance matrix, $\ell_{\textrm{mod}}$ (\emph{third column};
  marginal best-fit dipole axis (\emph{fourth column}) and amplitude
  (\emph{fifth column}) with 68\% confidence regions indicated;
  statistical significance of non-zero detection of $A$ (\emph{sixth
    column}); the change in maximum likelihood between modulated and
  isotropic models, $\Delta \log \mathcal{L} = \log \mathcal{L}_{\textrm{mod}}
  -\log \mathcal{L}_{\textrm{iso}} $ (\emph{seventh column}); and the Bayesian
  evidence difference, $\Delta \log E = \log E_{\textrm{mod}} - \log
  E_{\textrm{iso}}$ (\emph{eighth column}). The latter two were only
  computed for one data set, due to a high computational
  cost. However, other values can be estimated by comparing the
  significances indicated in the sixth column.}
\tablecolumns{8}
\tablehead{Data  & Mask & $\ell_{\textrm{mod}}$ &($l_{\textrm{bf}}, b_{\textrm{bf}}$) &
  $A_{\textrm{bf}}$ & Significance ($\sigma$) & $\Delta \log \mathcal{L}$ & $\Delta \log E$}
\startdata

ILC      & KQ85   & 64 & $(224^{\circ}, -22^{\circ}) \pm 24^{\circ}$ & $0.072 \pm 0.022$ & 3.3 & 7.3 & $2.6$  \\
$V$-band & KQ85   & 64 & $(232^{\circ}, -22^{\circ}) \pm 23^{\circ}$ & $0.080 \pm 0.021$ & 3.8 & \nodata & \nodata  \\
$V$-band & KQ85   & 40 & $(224^{\circ}, -22^{\circ}) \pm 24^{\circ}$ & $0.119 \pm 0.034$ & 3.5 & \nodata & \nodata  \\
$V$-band & KQ85   & 80 & $(235^{\circ}, -17^{\circ}) \pm 22^{\circ}$ & $0.070 \pm 0.019$ & 3.7 & \nodata & \nodata  \\
$W$-band & KQ85   & 64 & $(232^{\circ}, -22^{\circ}) \pm 24^{\circ}$ & $0.074 \pm 0.021$ & 3.5 & \nodata & \nodata  \\
ILC      & KQ85e  & 64 & $(215^{\circ}, -19^{\circ}) \pm 28^{\circ}$ & $0.066 \pm 0.025$ & 2.6 & \nodata & \nodata  \\
$Q$-band & KQ85e  & 64 & $(245^{\circ}, -21^{\circ}) \pm 23^{\circ}$ & $0.088 \pm 0.022$ & 3.9 & \nodata & \nodata  \\
$V$-band & KQ85e  & 64 & $(228^{\circ}, -18^{\circ}) \pm 28^{\circ}$ & $0.067 \pm 0.025$ & 2.7 & \nodata & \nodata  \\
$W$-band & KQ85e  & 64 & $(226^{\circ}, -19^{\circ}) \pm 31^{\circ}$ & $0.061 \pm 0.025$ & 2.5 & \nodata & \nodata \\  
ILC\tablenotemark{a}      & Kp2  & $\sim 40$ & $(225^{\circ},
-27^{\circ})\quad\quad\,\,\,\,$  & $0.11 \pm 0.04$ & 2.8 & 6.1 & $1.8$  
\enddata
\tablenotetext{a}{Results computed from $N_{\textrm{side}}=16$ and
  $9^{\circ}$ FWHM data, as presented by \citet{eriksen:2007b}.}
\end{deluxetable*}

\section{Data}
\label{sec:data}

In this paper we analyze several downgraded versions of the five-year
WMAP temperature sky maps, namely the template-corrected $Q$-, $V$-
and $W$-band maps, as well as the ``foreground cleaned'' Internal
Linear Combination (ILC) map \citep{gold:2009}. Each map is downgraded
to low resolution as follows \citep{eriksen:2007a}: First, each map is
downgraded to a HEALPix\footnote{http://healpix.jpl.nasa.gov}
resolution of $N_{\textrm{side}}=32$, by smoothing to an effective
resolution of $4.5^{\circ}$ FWHM and properly taking into account the
respective pixel windows. We then add uniform Gaussian noise of
$\sigma_{\textrm{n}} = 1\,\mu\textrm{K}$ RMS to each pixel, in order
to regularize the pixel-pixel covariance matrix at small angular
scales. The resulting maps have a signal-to-noise ratio of unity at
$\ell = 80$, and are strongly noise dominated at $\ell_{\textrm{max}}
= 95$. 

Two different sky cuts are used in the analyses, both of which are
based on the WMAP KQ85 mask \citep{gold:2009}. In the first case, we
directly downgrade the KQ85 cut to the appropriate
$N_{\textrm{side}}$, by excluding any HEALPix pixel for which more
than half of the corresponding sub-pixels are missing. This mask is
simply denoted KQ85. In the second case, we smooth the mask image
(consisting of 0's and 1's) with a beam of $4.5^{\circ}$ FWHM, and
reject all pixels with a value less than 0.99. We call this expanded
mask KQ85e. The two mask remove 16.3\% and 26.9\% of the pixels,
respectively.

The instrumental signal-to-noise ratio of the WMAP data is very high
at large angular scales, at about 150 for the V-band at
$\ell=100$. The only important noise contribution in the downgraded
sky maps is therefore the uniform regularization noise, which is not
subject to the additional beam smoothing. We therefore approximate the
noise covariance matrix by $N_{ij} = \sigma_{\textrm{n}}^2
\delta_{ij}$. Note that this approximation was explicitly validated by
\citet{eriksen:2007b} for the 3-year WMAP data, which have higher
instrumental noise than the 5-year data.

We also marginalize over a fixed set of ``foreground templates'',
$\mathbf{t}_i$, by adding an additional term to the data covariance
matrix on the form $\mathbf{F}_i =
\alpha_i\mathbf{t}_i\mathbf{t}_i^T$, with $\alpha_i \gtrsim 10^3$, for
each template. In addition to one monopole and three dipole
templates\footnote{ For an explicit demonstration of the importance of
  monopole and dipole marginalization on this specific problem, see
  Gordon 2007a}, we use the $V$--ILC difference map as a
template for both the $V$-band and ILC maps, the $Q$--ILC difference
for the $Q$-band, and the $W$--ILC difference for the
$W$-band. However, these foreground templates do not affect the
results noticeably in either case, due to the sky cuts used.

\section{Results}
\label{sec:results}

The main results from the analysis outlined above are summarized in
Table \ref{tab:results}. We consider nine different data combinations
(ie., frequency bands, masks and multipole range), and show 1) the
best-fit modulation axis and amplitude, both with 68\% confidence
regions; 2) the statistical significance of the corresponding
amplitude (ie., $A/\sigma_A$); and 3) the raw improvement in $\chi^2$
and Bayesian log-evidence for the modulated model over the isotropic
model. The last items are shown for the ILC with the KQ85 sky cut
only. For reference, we also quote the ILC result for the Kp2 mask
\citep{bennett:2003b} reported by \citet{eriksen:2007b} when analysing
the $N_{\textrm{side}}=16$ and $9^{\circ}$ FWHM data.

The reason for providing the full evidence for only one data set is
solely computational. The total CPU cost for the full set of
computations presented here was $\sim50\,000$ CPU hours, and the
evidence calculation constitutes a significant fraction of this. On
the other hand, the evidence is closely related to the significance
level $A/\sigma_A$, and one can therefore estimate the evidence level
for other cases in Table \ref{tab:results} given the two explicit
evidence values and significances. We have therefore chosen to spend
our available CPU time on more MCMC posterior analyses, rather than on
more evidence computations. 

We consider first the results for the ILC map with the KQ85 mask and
$\ell_{\textrm{mod}}=64$. In this case, the best-fit amplitude is $A =
0.073\pm0.022$, non-zero at the $3.3\sigma$ confidence level. The
best-fit axis points towards Galactic coordinates $(l,b) =
(224^{\circ}, -22^{\circ}$, with a 68\% uncertainty of
$24^{\circ}$. These results are consistent with the results presented
by \citet{eriksen:2007b}, who found an amplitude of $A=0.11\pm0.04$
and a best-fit axis of $(l,b) = (225^{\circ},-27^{\circ})$ for $\ell
\lesssim 40$.

\begin{figure}[t]

\mbox{\epsfig{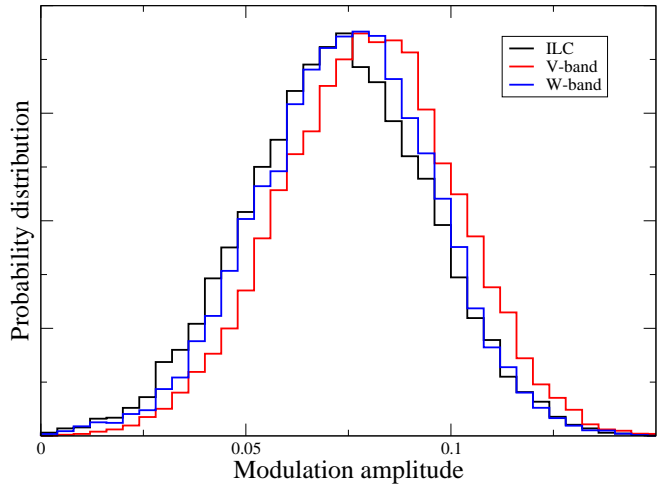}}

\caption{Posterior distributions for the dipole modulation amplitude,
  marginalized over direction and CMB power spectrum, computed for the
  KQ85 sky cut and $\ell_{\textrm{mod}}=64$.}
\label{fig:amplitude}
\end{figure}

\begin{figure}[t]

\mbox{\epsfig{figure=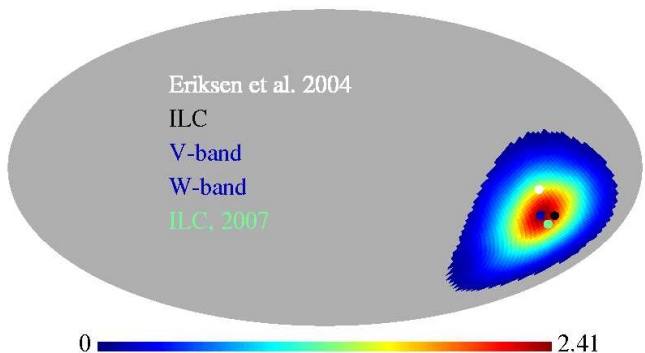,width=\linewidth,clip=}}

\caption{Posterior distribution for the dipole modulation
  axis, shown for the V-band map and KQ85 sky cut, marginalized
  over power spectrum and amplitude parameters. Grey sky pixels
  indicate pixels outside the $2\sigma$ confidence region. The dots
  indicate the axis 1) reported by \citet{eriksen:2004a} in white; 2)
  for both the ILC and V-band maps (these have the same best-fit axis)
  with the KQ85 sky cut in black; 3) for the $W$-bands in blue, and
  the axis reported by \citet{eriksen:2007b} in green. Note that the
  background distribution has been smoothed for plotting purposes to
  reduce visual Monte Carlo noise.}
\label{fig:axes}
\end{figure}

Second, we see that these results are only weakly dependent on
frequency, as both the V-band and W-band for the same mask and
$\ell$-range have amplitudes within $0.5\sigma$ of the ILC map, with
$A=0.080$ and $A=0.074$ and non-zero at $3.8\sigma$ and $3.5\sigma$,
respectively. (We have not included the Q-band analysis for the KQ85
mask, as there were clearly visible foreground residuals outside the
mask for this case.) The corresponding marginal posteriors are shown
in Figure \ref{fig:amplitude}, clearly demonstrating the consistency
between data sets. Figure \ref{fig:axes} compares the best-fit axes of
the three data sets, and also indicates the axes reported by
\citet{eriksen:2004a} and \citet{eriksen:2007b}.

Next, we also see that the results are not strongly dependent on the
choice of mask, as the amplitudes for the extended KQ85e mask are
consistent with the KQ85 results, even though it removes an additional
10\% of the sky. However, we do see, as expected, that the error bars
increase somewhat by removing the additional part of the sky, and this
reduces the absolute significances somewhat.

Finally, the best-fit modulation amplitudes for the V-band data and
KQ85 mask are $A=0.11$ for $\ell_{\textrm{mod}}=40$, $A=0.075$ for
$\ell_{\textrm{mod}} = 64$ and $A=0.066$ for $\ell_{\textrm{mod}}=80$
at $3.5\sigma$, $3.8\sigma$ and $3.7\sigma$, respectively. This is an
interesting observation for theoreticians who are interested in
constructing a fundamental model for the effect: Taken at face value,
these amplitudes could indicate a non-scale invariant behaviour of
$A$, as also noted by \citet{hansen:2008}. On the other hand, the
statistical significance of this statement is so far quite low, as a
single common value $A\sim0.07$ is also consistent with all
measurements. Better measurements at higher $\ell$'s are required to
unambiguously settle this question.

\section{Conclusions}
\label{sec:conclusions}

Shortly following the release of the first-year WMAP data in 2003,
\citet{eriksen:2004b} presented the early evidence for a dipolar
distribution of power in the CMB temperature anisotropy sky,
considering only the large angular scales of the WMAP data. Next,
\citet{groeneboom:2009} presented the evidence for a quadrupolar
distribution of CMB power, and found that this feature extended over
all $\ell$'s under consideration. Finally, \citet{hansen:2008} found
that the dipolar CMB power distribution is present also at high
$\ell$'s. The evidence for violation of statistical isotropy in the
CMB field is currently increasing rapidly, and the significance of
these detections are approaching $4\sigma$.

In this paper, we revisit the high-$\ell$ claims of
\citet{hansen:2008}, by applying an optimal Bayesian framework based
on a parametric modulated CMB model to the WMAP data at higher
multipoles than previously considered with this method, albeit lower
than those considered by \citet{hansen:2008}. In doing so, we find
results very consistent with those presented by \citet{hansen:2008}:
The evidence for a dipolar distribution of power in the WMAP data
increases with $\ell$. For example, when considering the V-band data
and KQ85 sky cut, the statistical significance of the modulated model
increases from $3.2\sigma$ at $\ell_{\textrm{mod}} =40$, to
$3.8\sigma$ at $\ell_{\textrm{mod}} = 64$, and $3.7\sigma$ at
$\ell_{\textrm{mod}}=80$.

The Bayesian evidence now also ranking within the ``strong to very
strong'' category on Jeffreys' scale. However, it should be noted that
the Bayesian evidence is by nature strongly prior dependent, and if we
had chosen a prior twice as large as the one actually used, the
corresponding log-evidence for the ILC map would have fallen from
$\Delta \ln E = 2.6$ to 1.7, ranking only as ``substantial''
evidence. For this reason, it is in many respects easier to attach a
firm statistical interpretation to the posterior distribution than the
Bayesian evidence.

It is interesting to note that the absolute amplitude $A$ may show
hints of decreasing with $\ell$. It is premature to say whether this
is due simply to a statistical fluctuation, or whether it might point
toward a non-scale invariant underlying physical effect, in which case
the amplitude $A$ should be replaced with a function $A(\ell)$. Either
case is currently allowed by the data.

To answer this question, and further constrain the overall model,
better algorithms are required. The current approach relies on
brute-force inversion of an $N_{\textrm{pix}} \times N_{\textrm{pix}}$
covariance matrix, and therefore scales as
$\mathcal{O}(N_{\textrm{pix}}^3)$ or
$\mathcal{O}(N_{\textrm{side}}^6)$. However, already the present
analysis, performed at $N_{\textrm{side}}=32$, required $\sim$50\,000
CPU hours, and increasing $N_{\textrm{side}}$ by an additional factor
of two would require $\sim$3 million CPU hours. More efficient
algorithms are clearly needed.

To summarize, there is currently substantial evidence for both dipolar
(Hansen et al.\ 2008 and this work) and quadrupolar power distribution
\citep{groeneboom:2009} in the WMAP data, and this is seen at all
probed scales. The magnitude of the dipolar mode is considerably
stronger than the quadrupolar mode, as a $\sim3.5\sigma$ significance
level is reached already at $\ell\sim 64$ for the dipole, while the
same significance was obtained at $\ell\sim 400$ for the quadrupole.

These observations may prove useful for theorists attempting to
construct alternative models for these features, either
phenomenological or fundamental. Considerable efforts have gone
towards this goal already \citep[e.g.,][]{ackerman:2007, boehmer:2008,
  carroll:2008a, carroll:2008b, chang:2008, erickcek:2008a, erickcek:2008b,
  gordon:2005, gum:2007, himmetoglu:2008a, himmetoglu:2008b,
  kahniashvili:2008, kanno:2008, koivisto:2008a, koivisto:2008b, pereira:2007,
  pitrou:2008, pullen:2007, watanabe:2009, yokoyama:2008}, but so far
no fully convincing model has been established. Clearly, more work is
needed on both the theoretical and observational side of this
issue. Fortunately, it is now only a few years until Planck will open
up a whole new window on these issues by producing high-sensitivity
maps of the CMB polarization, as well as measuring the temperature
fluctuations to arc-minute scales. We will then be able to measure the
properties of the dipole, quadrupole and, possibly, higher-order modes
of the modulation field to unprecedented accuracy.

\begin{acknowledgements}
  HKE acknowledges financial support from the Research Council of
  Norway. The computations presented in this paper were carried out on
  Titan, a cluster owned and maintained by the University of Oslo and
  NOTUR. Some of the results in this paper have been derived using the
  HEALPix \citep{gorski:2005} software and analysis package.  We
  acknowledge use of the Legacy Archive for Microwave Background Data
  Analysis (LAMBDA). Support for LAMBDA is provided by the NASA Office
  of Space Science.
\end{acknowledgements}

\end{document}